\documentclass{aa}  

\usepackage{graphicx}
\usepackage{txfonts}
\begin{document}

   \title{Formation of cold giant planets around late M dwarfs via core accretion and the fate of inner rocky worlds}
   \titlerunning{Cold giant planet formation around late M dwarfs}

   \subtitle{}

   \author{M. Sanchez \inst{1},
          N. van der Marel\inst{1},
          M. Lambrechts\inst{2},
          S. Paardekooper\inst{3},
          Y. Miguel\inst{1,4}
          }

   \institute{Leiden Observatory, Leiden University, P.O. Box 9513, 2300 RA Leiden,
The Netherlands.\\
\email{msanchez@strw.leidenuniv.nl}
\and
    Center for Star and Planet Formation, Globe Insitute, Øster Voldgade 5, 1350 Copenhagen, Denmark.
    \and
    Delft University of Technology, Postbus 5, 2600 AA Delft, The Netherlands.
    \and
    SRON Netherlands Institute for Space Research, Niels Bohrweg 4, 2333 CA Leiden, The Netherlands
    }

   \date{}

 
  \abstract
   {Modeling how cold giant planets form around M dwarfs remains a challenge, both because their protoplanetary disks can lack sufficient mass and because such planets are expected to migrate inward while interacting with the disk. Moreover, it remains unknown whether inner rocky planets can survive in systems that host a cold giant around very low-mass stars, which could have important implications for the habitability of rocky worlds.}
   {We investigated the conditions required for the formation of giant planets at large orbital distances (1–3 au) around a 0.1 M$_\odot$ star, and explored the circumstances under which a close-in rocky planet can survive.}
   {We performed N-body simulations in which planetary embryos grow through pebble accretion, followed by gas accretion during the disk lifetime. Assuming a local disk turbulent viscosity ($\alpha_{\rm{t}}$) of  $10^{-4}$, we included planet–disk interactions throughout the disk evolution, using a new prescription that accounts for the onset of outward migration when the planet-to-star mass ratio ($q$) exceeds $0.002$.}
   {We find that a cold giant planet can form around a late M dwarf, even with an initial pebble mass of only 6 M$_\oplus$, provided the disk gas mass is 10$\%$ of the stellar mass. This outcome requires a compact 20 au disk in which the inner, viscosity-dominated region has a high gas surface density set by a low accretion viscosity ($\alpha_{\rm g} = 10^{-4}$), that planet–planet collisions assemble a $\sim$5 M$_\oplus$ core within 1 Myr, and that the gas disk survives for 10 Myr. In addition, an inner rocky planet can survive in a close-in orbit if it migrates into the inner disk cavity before the outer body grows into a giant.}
   {The initial dust mass required for giant planet formation around very low-mass stars does not need to be as extreme as previously thought. A combination of planet–planet collisions, efficient pebble accretion, and a long disk lifetime plays a key role in enabling the formation of cold giant planets with masses between those of Saturn and Jupiter.}

               
   \maketitle

%

\section{Introduction}

Over the last decade, giant exoplanets have been discovered orbiting M dwarfs, and their number continues to grow \citep[e.g.,][]{GEMS2024}. Giants with orbital periods of less than 100 days are less common around M dwarfs compared to Sun-like stars, with occurrence rates below 1$\%$ \citep{Bryant2023}. In contrast, giants with orbital periods greater than 100 days show higher occurrence rates, ranging between 1$\%$ and 5$\%$ \citep{Clanton2014, Meyer2018}. In particular, all giant planets with masses above 0.3 M$_{\rm Jup}$ discovered around stars between 0.08 and 0.2 M$_\odot$ have orbital periods longer than 100 days and were detected mostly via microlensing\footnote{\url{https://exoplanetarchive.ipac.caltech.edu/}}. The orbital distances of single giant planets range from 0.4 to 5 au. From the observed sample, 19 correspond to single cold-giant-planet systems, representing about 85$\%$ of the population, while the remaining 15$\%$ consists of four planets belonging to two-giant-planet systems. Notably, no system has yet been found hosting both a giant planet and an inner rocky companion, unlike several systems around Sun-like stars \citep[e.g.,][]{Zhu2021}.

The mechanisms underlying giant planet formation remain poorly understood and continue to challenge existing theoretical models \citep[e.g.,][]{Liu2019,RBurn2021,Schlecker2022,Shibata2025}.  
 Limitations in dust and gas disk mass are thought to restrict the material available for core growth and gas accretion \citep[e.g.,][]{Ansdell2017, Pinilla2022}. For instance, the discovery of a close-in Neptune-mass planet orbiting a 0.1 M$_{\odot}$ star could only be explained with core accretion models that assume a disk mass an order of magnitude higher than typically observed \citep{Stefansson2023}. 
 Planet population synthesis models, when confronted with planets detected around M dwarfs by HARPS and CARMENES radial velocity surveys, failed to reproduce a population of giant planets around stars less massive than 0.5 M$_\odot$, suggesting a gap in our understanding of giant planet migration around very low-mass stars \citep{Schlecker2022}. Recently, \citet{Pan2024} demonstrated that giant planets can form around late M dwarfs with periods shorter than 100 days. In their models, lunar-mass protoplanets can grow through a combination of pebble accretion and planet–planet collisions provided that the disks are sufficiently massive, with $M_{\rm dust} > 50$ M$_\oplus$ and $M_{\rm gas} \sim 15$ M$_{\rm{Jup}}$, assuming a low gas surface density regulated by a high accretion viscosity ($\alpha_{\rm{g}}$) of $10^{-2}$ and a local turbulent viscosity ($\alpha_{\rm{t}}$) of $10^{-2}$-$10^{-3}$. However, observational studies suggest that turbulence in disks is better characterized by lower turbulent viscosities, closer to $10^{-4}$ \citep[e.g.,][]{Rosotti2023}. Moreover, the model assumed inward migration for giant planets, which may not be accurate for late M dwarfs. A new study based on hydrodynamical simulations using the \textit{FARGO3D} code, assuming high planet-to-star mass ratios in low-viscosity disks ($\alpha_{\rm{t}} = 10^{-4}$), shows that planets more massive than 66 M$_\oplus$ orbiting a 0.1 M$_\odot$ star can instead undergo outward migration \citep{Sanchez2025}.

This observational and theoretical context suggests that cold giant planets around the least massive stars likely require an alternative formation pathway. To address this, we performed a set of N-body simulations that implement the new prescriptions for giant planet migration around late M dwarfs proposed by \cite{Sanchez2025}, starting from planetary seeds that grow by pebble and gas accretion. Our findings point to a new pathway for the formation of cold giant planets around the least massive M dwarfs and the survival of inner rocky worlds.

\section{Planet formation model}

We used a modified version of the $N$-body code \textsc{Mercury} \citep{Chambers1999} that includes additional physical processes: 
\begin{itemize}
\item An evolving gas-disk model. We adopted the disk model proposed by \citet{Ida2016}, which describes a viscous disk in the inner region and an irradiated disk in the outer region. 
We assumed an initial disk age of 1 Myr and a disk lifetime of 10 Myr, consistent with disk observations of M dwarfs \citep[e.g.,][]{Downes2015,Pfalzner2022}.
\item Star--planet tidal interactions and general relativistic corrections. We followed the equilibrium tidal model adapted by \cite{Bolmont2011} for the case of planets around brown dwarfs and M dwarfs. We included acceleration corrections due to relativistic effects proposed by \cite{Anderson1975}. 
\item Planet–disk interactions for rocky planets. We implemented the dynamical friction approach proposed by \citet{Ida2020} for the acceleration corrections experienced by planets embedded in the disk, and the non-isothermal torque prescription of \citet{Paardekooper2010,Paardekooer2011}. 
\item Growth by pebble accretion. We let planets grow via pebble accretion until they reach the isolation mass \citep{Bitsch2018}, following the model of \citet{LJM2014}, with efficiencies for planets on eccentric orbits given by \citet{OL2018} and \citet{LO2018}, assuming that the planet’s inclination satisfies $i<h_{\rm peb}$, with $h_{\rm peb}$ the pebble aspect ratio. 
\end{itemize}
The detailed prescriptions of these effects are presented in \citet{Sanchez2020,Sanchez2022,Sanchez2024}. In this work, we present an updated version of the code that also includes the following modifications (see the appendix for details):
\begin{itemize}
    \item New parameterization of the pebble flux. We defined an evolving pebble flux set by a squared exponential decay to ensure consistency with the dust mass evolution inferred from observations \citep{Appelgren2023}, expressed as
\begin{align}
M_{\mathrm{flux}} = M_{\mathrm{flux,in}} \exp\left(-t / Myr\right)^{2},
\end{align}
with $M_{\mathrm{flux,in}}= 2.5 \times 10^{-5}$~M$_\oplus yr^{-1}$ the initial pebble flux at 1 Myr, set to half of the flux assumed in \cite{Sanchez2024}. We assumed a compact disk with a radius of 20 au and a pebble density profile consistent with millimeter-flux observations \citep{Sanchez2024,Osmar2025}.
    \item New inputs for star--planet tidal interactions. We included the necessary inputs to also account for the interaction of giant planets around M dwarfs \citep{Bolmont2015}. 
    \item Gas accretion onto the planets. We let planets accrete gas once they reach the isolation mass, as in \cite{Pan2024}. 
    We highlight that the mass needed for a planet to achieve an accretion rate of $10^{-5}$~M$_\oplus\mathrm{yr}^{-1}$ is $\sim 5$~M$_\oplus$. At this rate, the planet can double its mass within $\sim 0.5$ Myr, thereby triggering the onset of runaway gas accretion.
     \item Updated planet–disk interactions. We implemented updated prescriptions to calculate the torques exerted by the disk onto the planets. For gap-opening planets with planet-to-star mass ratios ($q$) satisfying $q < 0.002$ 
     we adopted the prescription of \citet{Kanagawa2018}, while for planets with $q > 0.002$, we applied the new torque prescription proposed by \cite{Sanchez2025}, which accounts for outward migration.
\end{itemize}

We characterized the code following \citet{Sanchez2024}, adopting the hybrid integrator and modeling collisions between bodies as perfect mergers. We ran a set of ten simulations for 50 Myr, with a central object of 0.1 M$_\odot$ and 25 lunar-mass planetary seeds, a mass value commonly adopted in previous formation models around M dwarfs \citep[e.g.,][]{Coleman2019, Sanchez2024, Pan2024}. We note that even in the inner disk, a seed with only 1$\%$ of the lunar mass can still undergo efficient pebble accretion \citep{OL2018, LO2018}. The seeds are initially distributed between 0.1 and 2 au, with separations of 15–30 mutual Hill radii, slightly larger than those typically used \citep[e.g.,][]{Kokubo2000}. Previous simulations with more compact configurations and larger samples \citep{Sanchez2024} showed no significant differences in the early system evolution compared to this work. We used a two-$\alpha$ model: one, $\alpha_{\rm g}$, to characterize the global angular momentum transport that regulates the gas surface density evolution, and another, $\alpha_{\rm t}$, associated with local disk turbulence, which controls the vertical stirring of pebbles and planet–disk interactions. For our main set of simulations, we adopted $\alpha_{\rm {t}} = \alpha_{\rm g} = 10^{-4}$. The initial pebble mass ($M_\mathrm{peb}$) was $6$~M$_\oplus$ and the gas mass ($M_{\rm gas}$) was $10$~M$_{\rm Jup}$, corresponding to 10$\%$ of the stellar mass. These values were obtained by integrating the initial pebble surface density profile and the gas surface density profile, respectively, over the disk radius of 20 au (see profiles in \citealt{Sanchez2024}). The resulting pebble-to-gas ratio is $10^{-3}$, which represents a limiting case since not all dust is assumed to be in pebble form \citep[e.g.,][]{LJ2014}. The Stokes numbers within 2 au range between 0.02 and 0.05, corresponding to centimeter-sized particles at the midplane due to the high gas density. This remains fully consistent with ALMA observations since in dense disks, a vertically settled and coagulated midplane population of centimeter-sized pebbles can still be entirely compatible with the observed millimeter-sized emission. This is because the observed flux depends primarily on the integrated dust mass rather than on the local midplane size distribution. 
To explore the effect of lower disk masses, we used the same initial conditions of the simulations that formed a giant planet but reduced the gas masses to $0.05~M_\star$ and $0.01~M_\star$, by lowering the gas disk surface density profiles assuming $\alpha_{\rm{g}}=10^{-3}$ and $\alpha_{\rm{g}}=10^{-2}$, respectively. Additionally, we extended the disk lifetime to 15 Myr to assess its impact on the final mass that the giant planet could reach.

\section{From seed to cold giant}

 We investigated the conditions required for a planetary seed to grow into a cold giant planet more massive than 0.3 M$_{\rm{Jup}}$ around a star of 0.1 M$_\odot$ using our numerical model. Here we describe the dynamical evolution of the cold giant planet within the system in which it is formed, and the impact on the survival of an inner rocky planet companion.

\subsection{Conditions needed for the formation of a cold giant planet}

\begin{figure}
    \centering
    \includegraphics[width=0.9\linewidth]{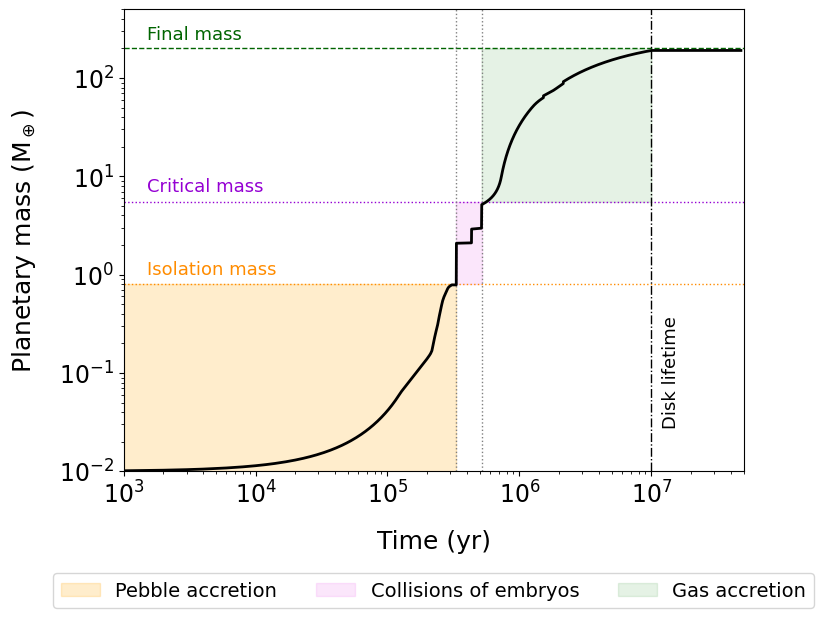}
    
    \includegraphics[width=0.9\linewidth]{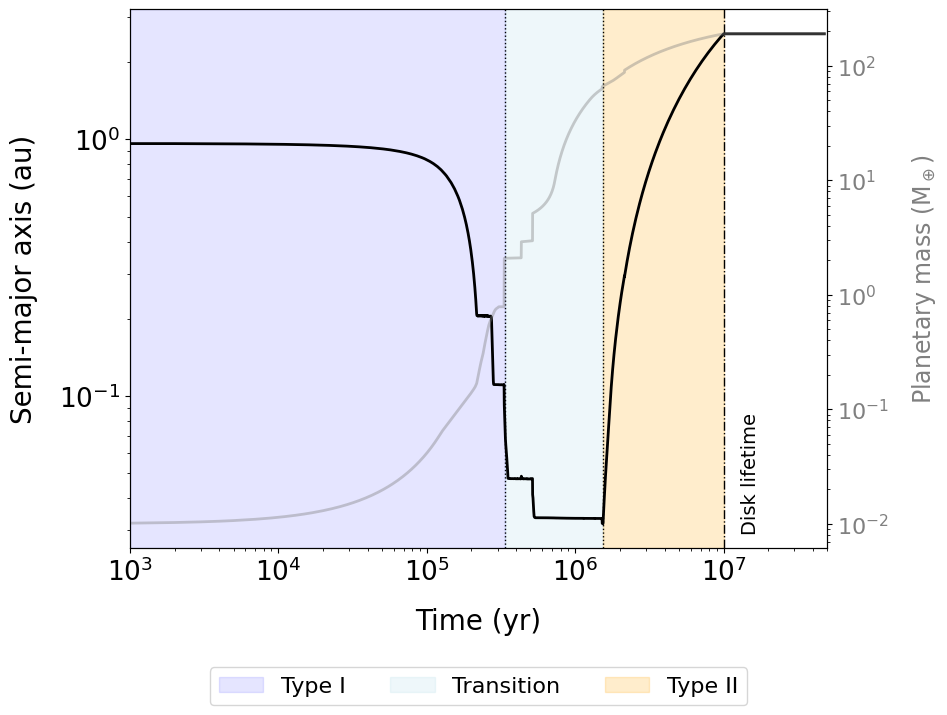}    
    \caption{Evolutionary pathway of cold giant planet formation. Top: Evolution of the planetary mass (black line) across different growth regimes: the pebble-accretion–dominated phase (orange area), limited by the isolation mass (dotted orange line); the phase dominated by planet–planet collisions  (pink area), with the critical mass required to trigger runaway gas accretion indicated (dotted pink line); and the gas-accretion–dominated phase (green area), during which the planet grows until it reaches its final mass (dotted green line) before disk dispersal (dash-dotted black line). Bottom: Evolution of the semi-major axis (black line) across the different migration regimes: type I migration (light blue area), transition regime where migration slows (gray area), and outward migration (orange area).}
    \label{fig:criticalplots}
\end{figure}

 The evolution of the mass and semi-major axis of a planetary seed that becomes a cold giant planet is shown in Fig. \ref{fig:criticalplots}. In the early stages, growth is dominated by pebble accretion, until the outermost planet reaches the isolation mass within $\sim$300,000 yr after migrating inward on a short timescale. Thereafter, collisions between Earth-mass embryos drive further growth, allowing the planet to reach $\sim$5 M$_\oplus$. Within the following million year, the planet's mass grows to $\sim$10 M$_\oplus$, the critical mass required to trigger runaway gas accretion. Subsequently, the planet grows to $\sim$66 M$_\oplus$ within $\sim$2 Myr, with its growth regulated by the gas supply set by the stellar accretion rate. Once the planet reaches this mass, it transitions to a different migration regime (see the appendix for details), reverses its migration direction, and begins to migrate outward. In the late stages of disk evolution, the planet accretes in the Hill regime, but once it migrates beyond $\sim$1 au, the stellar accretion rate again limits the gas supply, slowing its growth until disk dispersal. The final mass of the planet depends on how quickly it reaches the $\sim$5 M$_\oplus$ threshold. Including gas accretion at such low core masses is essential rather than waiting until $\sim$10M$_\oplus$ as in classical models. Notably, the 5 M$_\oplus$ threshold is consistent with the minimum mass required for nucleated instability to begin \citep{Rafiknov2006}. Outward migration is crucial in this scenario, as without it giant planets could not reach wide orbits. Pebble accretion efficiencies are high ($\sim50\%$) because seeds grow mainly within 0.5~au. While we do not include pebble fragmentation inside the snow line \citep[e.g.,][]{Venturini2020}, which could reduce accretion, most inner seeds either survive as one close-in planet or collide with the star. In contrast, giant planets form from seeds beyond the snow line. Thus, including fragmentation would likely have little impact on their formation but might require a higher dust mass to reach similar growth.

\subsection{Inner rocky planet survival in a system with a cold giant}

\begin{figure*}
    \centering
    \includegraphics[width=0.24\linewidth]{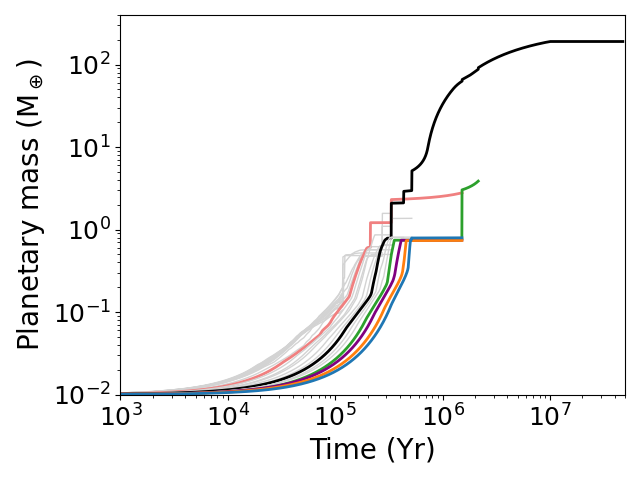}
     \includegraphics[width=0.24\linewidth]{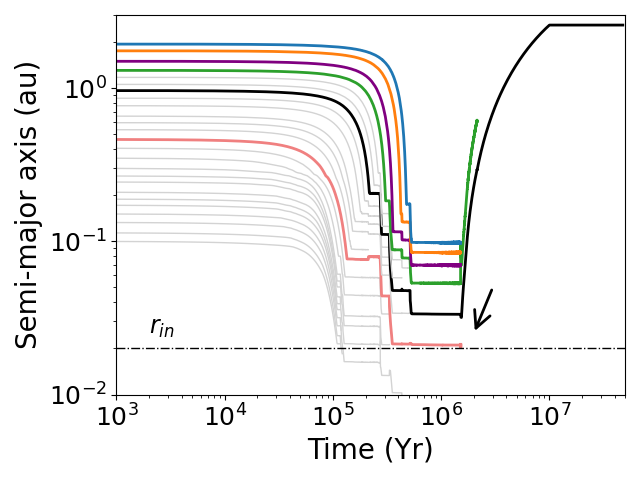}
      \includegraphics[width=0.245\linewidth]{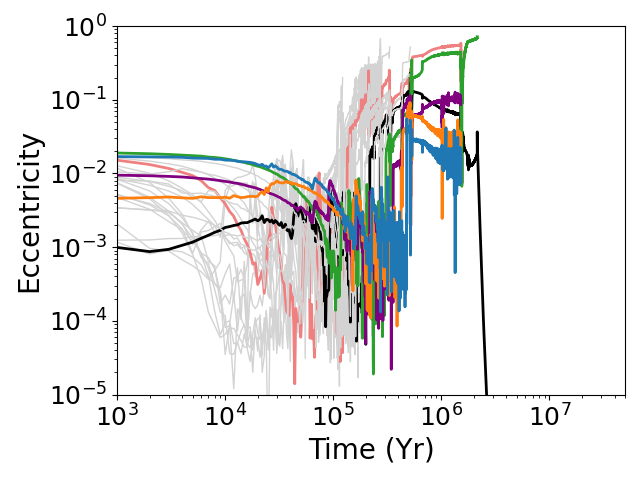}
    \includegraphics[width=0.24\linewidth]{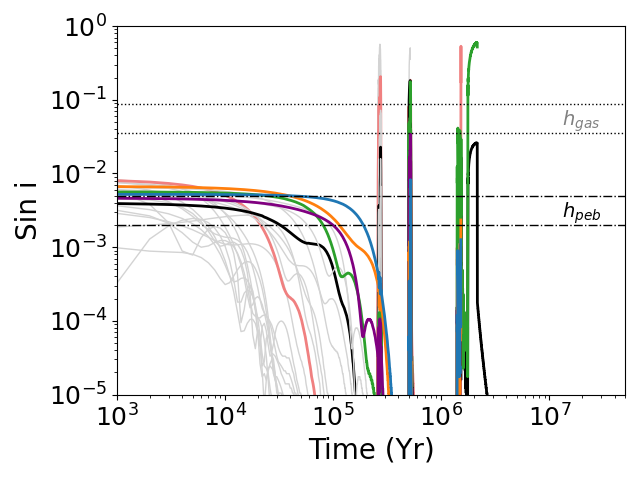} 
    
      \includegraphics[width=0.24\linewidth]{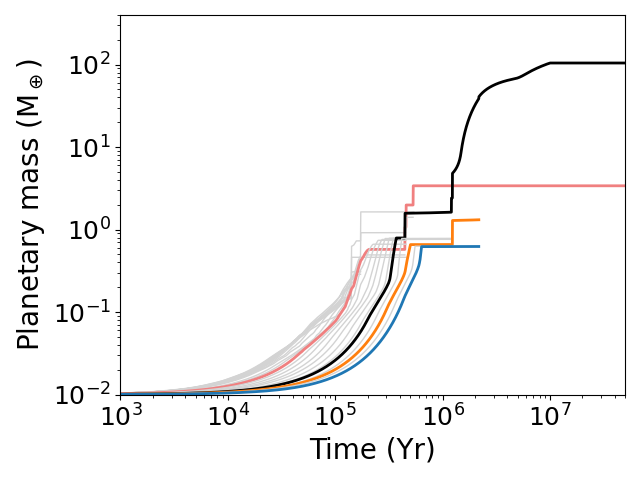}
     \includegraphics[width=0.24\linewidth]{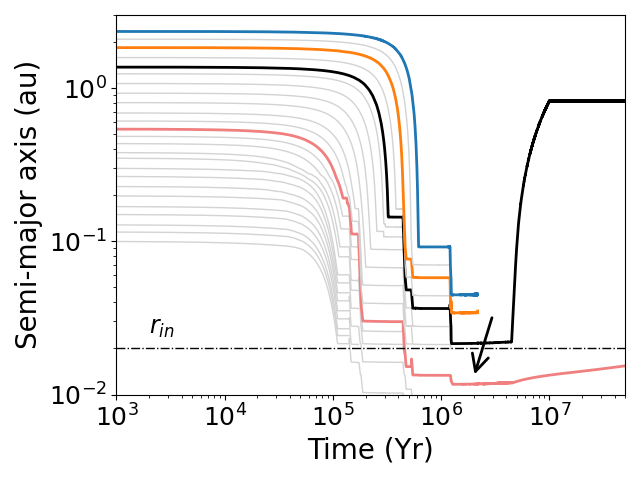}
      \includegraphics[width=0.245\linewidth]{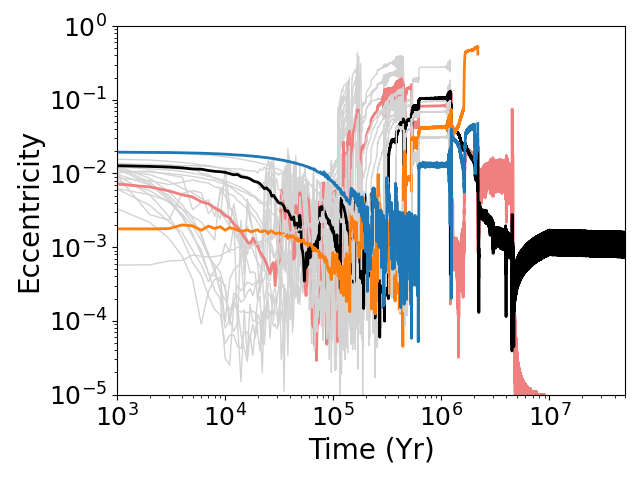}
      \includegraphics[width=0.24\linewidth]{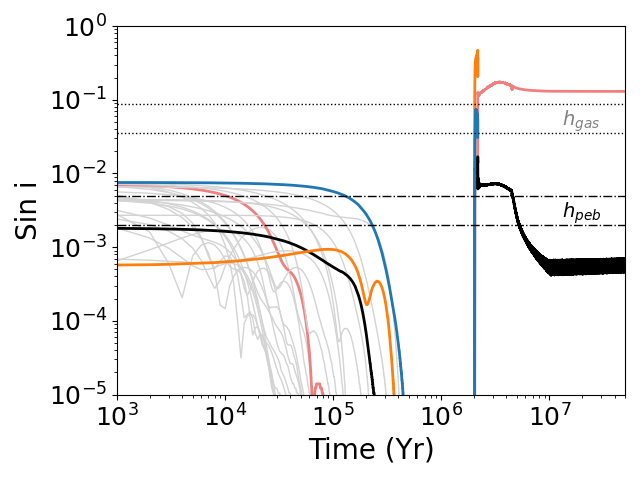}
    \caption{Evolution of the planetary mass, semi-major axis, eccentricity, and inclination of the planetary seeds (gray lines) in systems where a single giant planet forms (top row), and where a close-in rocky planet forms together with a cold giant planet (bottom row). We show the evolution of the giant planet (black line) and the evolutions the other planets that survived in the system (colored lines) before the more massive planet became a giant planet. The inner edge of the disk, $r_{\rm{in}}$, is indicated (horizontal dashed black line). The location of the innermost planet (pink line) is marked with an arrow for comparison with the inner edge of the disk. The gas aspect ratio, $h_{\rm{gas}}$ (dotted black lines), and the pebble aspect ratio, $h_{\rm{peb}}$ (dotted-dashed black lines), are also overplotted, evaluated at $a=0.02$~au and $a=2$~au.}
    \label{fig:evolutiontrend}
\end{figure*}

We analyzed the evolutionary pathways of two simulations in which a giant planet was formed: one leading to a system with a single cold giant, and the other to a system hosting both a cold giant and a close-in rocky companion. Figure \ref{fig:evolutiontrend} shows the evolution of the planetary mass, $M_{\rm{p}}$, semi-major axis, $a$, eccentricity, $e$, and  inclination, $i$, for both cases. In the first 0.5 Myr, planet growth is dominated by pebble accretion, and shortly thereafter by collisions among embryos. By 1 Myr, both systems consist of several Earth-like planets and one protoplanet that reaches $\sim$5 M$_\oplus$, triggering rapid gas accretion and the formation of a giant planet. As the giant grows, it accretes the outer planets, either while still migrating inward (two-planet case) or immediately after beginning outward migration (single-planet case). In the single-planet system, the innermost planet is also accreted by the giant. In contrast, in the two-planet system the innermost planet survives because it enters the inner disk cavity early in its evolution and settles into a stable orbit at $\sim$0.015 au.
By the end of the disk’s lifetime, the giant in the single-planet system has a mass of $\sim$0.6 M$_{\rm Jup}$ at 2.6 au, while in the two-planet system the giant's mass is $\sim$0.3 M$_{\rm Jup}$, as it crosses the 5 M$_\oplus$ threshold later in its evolution. In both scenarios, the surviving planets remain on quasi-circular orbits: the innermost planet due to strong star–planet tidal interactions, and the giant planet due to eccentricity damping during the disk lifetime. During the pebble accretion phase, most planets satisfy $i < h_{\rm peb}$, and for most of the integration time $i < h_{\rm g}$, where $h_{\rm peb}$ and $h_{\rm g}$ are the pebble and gas aspect ratios, respectively \citep[see][]{Sanchez2024}. Close encounters can temporarily raise the inclinations up to $\sim 5h_{\rm g}$. In both systems, the giant planet remains near the midplane, while in the system with an inner rocky companion, the rocky planet retains a high inclination due to its location inside the disk cavity.

In our explored scenarios, the survival of an inner rocky planet in systems with a cold giant primarily depends on whether it enters the inner disk early, preventing collisions with the growing giant and allowing coexistence. Under these conditions, the rocky planet remains atmosphere-free, but further simulations and longer integrations are needed to explore its composition and long-term evolution, which lie beyond the scope of this work.

\section{Discussions}

In our simulations, giant planet formation occurs in low-viscosity environments characterized by 
$\alpha_{\rm t}=10^{-4}$. This appears to differ from the findings of \citet{Pan2024}, who identify giant planet formation as most favorable for higher turbulent viscosities, $\alpha_{\rm t}=10^{-3}$. However, this distinction arises primarily from the different gas surface density profiles adopted in the two studies. Although \citet{Pan2024} assumed a more massive disk (15$\%$ of the stellar mass, compared to our 10$\%$), their choice of a higher accretion viscosity of $10^{-2}$ yields a significantly lower gas surface density than our simulations. In contrast, our model uses $\alpha_{\rm g}=10^{-4}$, resulting in gas densities that are a factor 30 higher than those in \citet{Pan2024}. This difference in disk structure has direct dynamical consequences. In our denser disk, type I migration is substantially faster, up to two orders of magnitude, than in lower-density disks, causing embryos to converge toward the inner disk on timescales of less than 0.1–0.2 Myr (see Fig. \ref{fig:evolutiontrend}), compared with the 1–2 Myr in \citet{Pan2024}. Such rapid migration strongly increases mutual collision rates, allowing planetary cores of 5 M$_\oplus$ to form very early (within 0.5 Myr). Provided that the disk lifetime is sufficiently long (up to 10 Myr), these cores can subsequently undergo gas accretion and grow to Saturn–Jupiter masses.

We also note a key difference in the pebble reservoir required to form giant planets. In \citet{Pan2024}, giant planet formation required pebble masses greater than 50 M$_\oplus$, whereas our model begins with a much smaller reservoir of 6 M$_\oplus$ (total of 10 M$_\oplus$). This could be related to the fact that their slower-migration environment produces fewer embryo–embryo collisions, making growth more dependent on reaching the pebble isolation mass, a process less efficient at large orbital distances. In contrast, the much faster migration in our dense disk makes early collisional growth the dominant pathway, allowing some embryos to reach the critical core mass even with a modest pebble reservoir. This strong dependence on planet–planet collisions also explains why only a subset of our simulations produce cores massive enough to trigger runaway gas accretion. Overall, the high gas surface density in our model plays a central role by accelerating migration, increasing early dynamical interactions, and enabling the rapid formation of massive planetary cores in disks with relatively small pebble reservoirs.

The pebble accretion efficiencies in our main simulations did not include the inclination-dependent reduction factor. However, to assess its impact, we repeated two runs that formed a giant planet, now incorporating the efficiency reduction for cases where 
$i>h_{\rm peb}$, as in \citet{OL2018}. As expected, pebble accretion efficiency decreased by about a factor of 2 during the first $10^{5}$ yr, when some embryos briefly reached inclinations comparable to the pebble aspect ratio. However, by $3\times10^{5}$ yr, a similar number of embryos reached the pebble isolation mass in the two sets of simulations. This is because planet–planet collisions consistently occurred after 
$(1.4\pm 0.05)\times10^{5}$
 yr in all runs, demonstrating that collisional growth is the dominant mechanism driving embryos in the inner disk to reach isolation masses. Therefore, neglecting the early-time reduction does not alter when or where embryos reach this critical stage.
Regarding giant planet formation, one revised simulation still produced a giant planet, while the other did not. This divergence did not arise from differences in pebble accretion, but rather from stochastic variations in the dynamical histories, which in one case prevented the sequence of collisions needed to assemble a sufficiently massive core. This underscores that, in our model, collisional growth and the system’s intrinsic dynamical sensitivity primarily determine whether giant planet formation is ultimately achieved.

\section{Conclusions}

We performed N-body simulations to explore the formation of cold giant planets around a 0.1~M$_\odot$ star in a low-viscosity disk ($\alpha_{\rm t}=10^{-4}$). We find that giant planet formation does not require extreme disk conditions in low-viscosity environments where pebble drift is efficient. For instance:
\begin{itemize}
    \item Saturn- and Jupiter-like planets can form in disks with a pebble supply sustained over $\sim 3 \times 10^5$ yr and a total pebble mass of $\sim 10$ M$
 _\oplus$. The total disk mass should be $\sim 10\%$ of the stellar mass for a compact disk with a high gas surface density (20 au, $\alpha_{\rm g}=10^{-4}$). 
 \item Additional simulations suggest that cold-giant formation is unlikely in disks less massive than 5 $\%$ of the stellar mass, which are associated with lower gas density profiles and higher accretion viscosities ($\alpha_{\rm g}$  $> 10^{-4}$).
 \item A combination of planet–planet collisions and efficient pebble accretion before 1 Myr, together with a long disk lifetime of at least 10 Myr, plays a key role in enabling the formation of a cold giant planet. A long disk lifetime has also been found to be crucial for giant planet formation in recent studies \citep[e.g.,][]{Shibata2025}.  However, extending the disk lifetime to 15 Myr results in a 5$\%$ difference in the final planetary mass, indicating that extending the disk lifetime to more than 10 Myr has a negligible impact on the giant planet formation process. 
\end{itemize} 

We note that three of our ten simulations produced a giant planet (two with a single giant and one with a giant and an inner rocky companion), while the others formed Earth- and super-Earth-like planets. A larger set of simulations, currently in progress, will better capture the low occurrence rate of cold giants and further clarify how different initial pebble masses and gas density profiles shape planetary diversity.

\begin{acknowledgements}
    This work was performed using the compute resources from
the Academic Leiden Interdisciplinary Cluster Environment (ALICE) provided
by Leiden University. This work made use of the NASA Exoplanet Archive,
which is operated by the California Institute of Technology, under contract with
the National Aeronautics and Space Administration under the Exoplanet Exploration Program. The authors thank Gudmundur Stefánsson for his helpful comments, and the referee for their detailed report, which has improved the clarity of our manuscript.
\end{acknowledgements}

%
%
\bibliographystyle{aa} 
\bibliography{biblioUp} 

\begin{appendix} 
\section{Model of gas accretion onto planets}
\begin{figure}
    \centering    \includegraphics[width=0.9\linewidth]{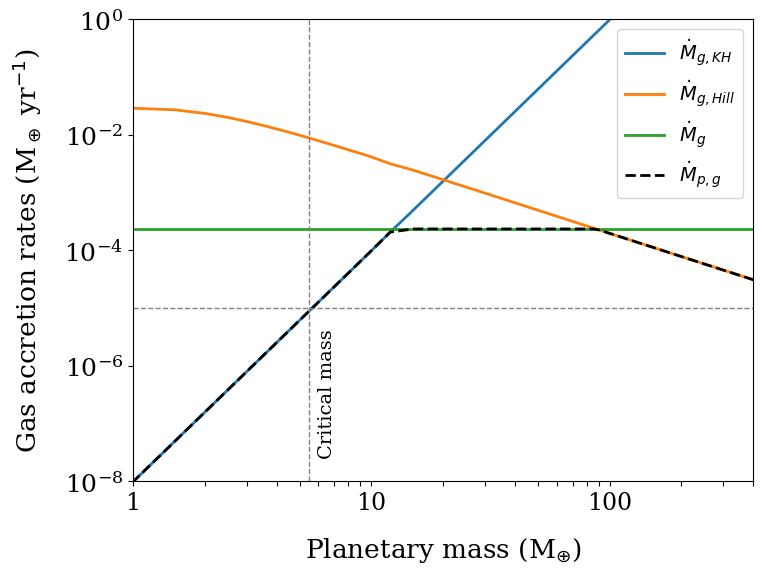}
    \caption{Gas accretion rates as a function of planetary mass. The curves show the Kelvin–Helmholtz contraction regime (blue line), the Hill regime (orange line), and disk-limited accretion (green line). The effective accretion rate onto the planet, defined as the minimum of the three regimes, is shown (dashed black line). The critical planetary masses at which the accretion rate reaches $10^{-5}$~M$_\oplus$yr$^{-1}$ are highlighted (dotted gray lines).}
    \label{fig:gasaccrates}
\end{figure}
We modified the core of  Mercury N-body code \cite{Chambers1999} to include gas accretion onto the planets. We assumed gas accretion starts once the planets reached the pebble isolation mass  \citep{Ogihara2020}. The planets accrete with a gas accretion rate defined as
\begin{equation}
\dot{M}_{\rm{p,g}} = \min \left(\dot{M}_{\rm{g,KH}} , \dot{M}_{\rm{g,Hill}}, \dot{M}_{\rm{g}} \right)
,\end{equation}
with $\dot{M}_{\rm{g,KH}}$ the accretion rate following Kelvin-Helmholtz contraction, which is expected to start after the planet reaches the isolation mass \citep{Ikoma2000} is set as follows:
 \begin{equation}
\dot{M}_{\rm{g,KH}} = 8 \times 10^{-8} \left(\frac{M_p}{3\,M_\oplus}\right)^4 \left(\frac{\kappa_{\mathrm{env}}}{1\,\mathrm{cm}^2\,\mathrm{g}^{-1}}\right)^{-1} \mathrm{M_\oplus\,\mathrm{yr}^{-1}}
,\end{equation}
with $M_{\rm{p}}$ the mass of the planet and $\kappa_{\rm{env}}$ the envelope opacity, that as it needs to be lower than the disk opacity (in our disk model is set at 1 $cm^{2}yr^{-1}$) we set at 0.1 $cm^{2}yr^{-1}$ as in \cite{Pan2024}.
As the planet increases it mass it will change the accretion rate regime. It could start accreting in the Hill regime, where only a fraction of gas within the planet Hill sphere
can be accreted \citep{Tanigawa2002}. Following the expression as appears in \cite{Liu2019} the accretion rate in the Hill regime can be express as follows: 
 \begin{align}
    \dot{M}_{\rm{g,Hill}} &= 0.004 \left(\frac{M_p}{3\,M_\oplus}\right)^{2/3} \left(\frac{M_\star}{0.1\,M_\odot}\right)^{-2/3} \left(\frac{\dot{M}_g}{10^{-8}\,M_\odot\,\mathrm{yr}^{-1}}\right) \times \nonumber \\
&\quad \times \left(\frac{\alpha_g}{10^{-2}}\right)^{-1} \left(\frac{h_g}{0.065}\right)^{-2} 
\left[1 + \left(\frac{M_p}{M_{\mathrm{gap}}}\right)^2 \right]^{-1} \mathrm{M_\oplus\,\mathrm{yr}^{-1},}
\end{align}
with $\alpha_{\rm{g}}$ the viscosity related to gas accretion, $h_{\rm{g}}$ the height scale of the gas disk (see the disk model in \citealt{Sanchez2022,Sanchez2024}), and $\dot{M}_{\rm{g}}$ the disk accretion rate which in this work is given by a fitting with observations in the Orion Nebular Cluster made by \cite{Manara2012}, as used to describe our disk model (see the appendix in \citealt{Sanchez2024}) which is express as follows:
\begin{equation}
\begin{split}
    \log\left(\frac{\dot M_{\textrm{g}}}{\textrm{M}_\odot~ \textrm{yr}^{-1}}\right)=&-5.12-0.46\log\left(\frac{t}{\textrm{yr}}\right)-5.75\log\left(\frac{M_{\star}}{\textrm{M}_{\odot}}\right)\\
    &+1.17\log\left(\frac{t}{\textrm{yr}}\right)\log\left(\frac{M_{\star}}{\textrm{M}_{\odot}}\right).
\label{eq:accretion_rate}
\end{split}    
\end{equation}
We estimate the necessary mass for a planet to open a gap in the disk following \cite{Kanagawa2015}:
\begin{equation}
    \begin{split}
       M_{\rm{gap}}= 17~\left(\frac{1}{\Sigma_{\rm{p}}/\Sigma_0}-1\right)^{0.5} \left(\frac{h_{\rm{g}}}{0.1}\right)^{2.5}\left(\frac{\alpha_{\rm{t}}}{10^{-3}}\right)^{0.5}\left(\frac{M_\star}{0.1 M_\odot}\right) ~~~\mathrm{M_\oplus}
    \end{split}
    \label{eq:Mgap}
,\end{equation}
with $\Sigma_{\rm{p}}/\Sigma_0$ the gas depth associated with gap-opening planets, which we set at 0.08 to assure that it will always be higher that the pebble isolation mass assumed in this work that follows \citet[see details in \citealt{Sanchez2024}]{Bitsch2018}.

As an example, in Fig. \ref{fig:gasaccrates} we show the accretion rates in the different regimes described above for planets with different masses around a 0.1 M$_\odot$ star, located close to the inner edge of the disk, at 0.02 au. We assumed a gas-disk of 1 Myr (see the gas-disk model in \citealt{Sanchez2024}). We highlight the critical mass of around 5 M$_\oplus$ associated with the accretion rate of $10^{-5}$~M$_\oplus\mathrm{yr}^{-1}$, which can trigger the runaway gas accretion.

\section{Planet--disk interactions}

\begin{figure}
    \centering
    \includegraphics[width=0.95\linewidth]{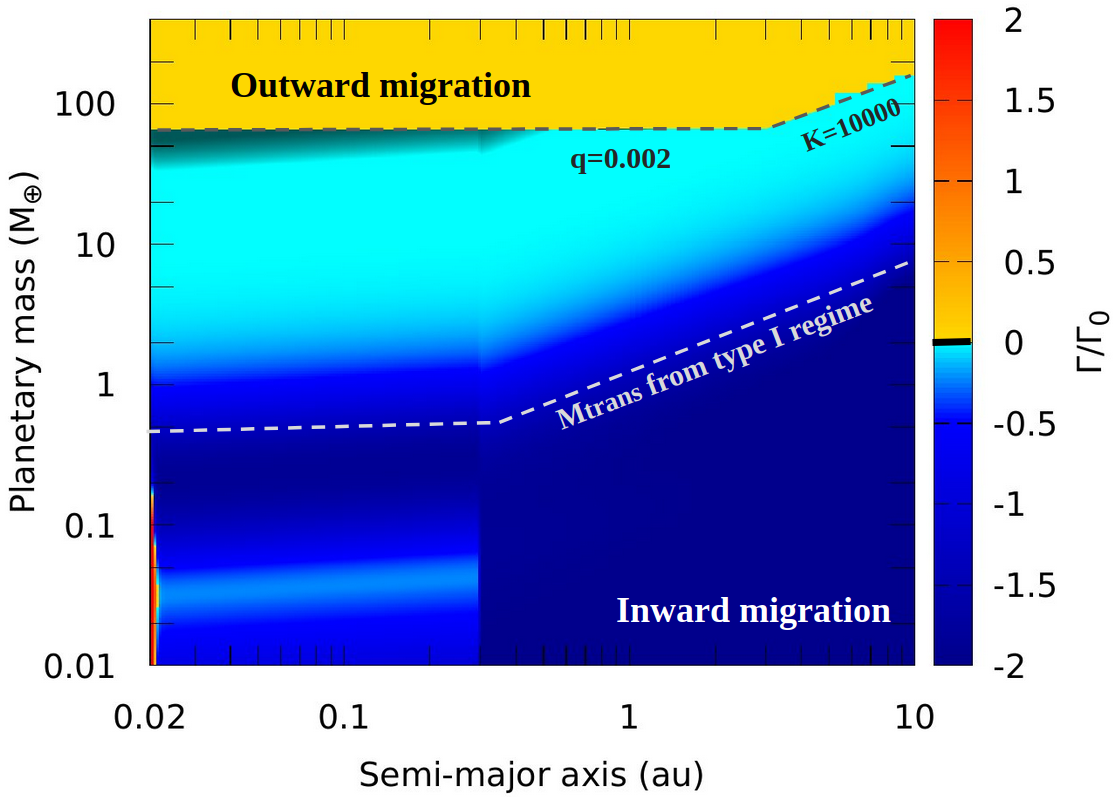}
    \caption{Normalized torques as a function of planetary mass and semi-major axis, assuming circular and coplanar orbits. The different migration regimes are indicated, separating inward from outward migration. The mass required to open a gap in the disk (dashed white line), the planet-to-star mass ratio ($q$) of $0.002$, and the gas depth criterion $K=10^{4}$ are over-plotted  (dashed black lines).}
    \label{fig:torques}
\end{figure}

We implemented an updated routine to account for the torques exerted by the gas disk on planets across different regimes. For all planetary masses, we retained the same expressions for the acceleration corrections already included in the code \citep{Sanchez2022}, adopting the formulation proposed by \citet{Ida2020}:
\begin{equation}
    \textbf{f}_{\textrm{gas}}=-\frac{v_\textrm{r}}{t_\textrm{e}}\hat{e_\textrm{r}}-\frac{(v_\theta-v_\textrm{k})}{t_\textrm{e}}\hat{e_\theta}-\frac{v_\textrm{k}}{2t_\textrm{a}}\hat{e_\theta}-\frac{v_\textrm{z}}{t_\textrm{i}}\hat{e_\textrm{z}},
\end{equation}
with the velocity of the embryos given by $\vec{v}=(v_\textrm{r},v_\theta,v_\textrm{z})$, and  $\hat{e_\textrm{r}}$, $\hat{e_{\theta}}$ and $\hat{e_\textrm{z}}$ the versors in each direction associated with a system in cylindrical coordinates $(r,\theta,z)$. The gas velocity is given by $\vec{v_\textrm{g}}=(0,(1-\eta)v_{\textrm{k}},0)$, with $ \eta \sim h_{\rm{g}}^{2}$ and $v_\textrm{k}$ the Keplerian velocity. The variables $t_\textrm{a}$, $t_\textrm{e}$ and $t_\textrm{i}$ denote the characteristic timescales for the orbital evolution of the semi-major axis $a$, eccentricity $e$, and inclination $i$ of the embryos, respectively. In all cases, we calculate $t_{\rm{a}}$ as follows:
\begin{equation}
    t_{\textrm{a}} = -\frac{t_{\textrm{wave}}}{2h_{\textrm{g}}^{2}}\frac{\Gamma_0}{\Gamma_\textrm{total}},
\end{equation}
with 
\begin{equation}
t_{\textrm{wave}} = \left(\frac{M_{\star}}{M_{\textrm{p}}}\right)\left(\frac{M_{\star}}{\Sigma_{\textrm{g}}r^{2}}\right)h_{\textrm{g}}^{4}\Omega_{\textrm{k}}^{-1},
\label{eq:twave}
\end{equation}
the timescale proposed by \citet{Papaloizou2000} and \citet{Tanaka2004}, and the scaling torque factor:
\begin{equation}
\Gamma_0=\left(M_\textrm{p}/M_\star\right)^2\Sigma_\textrm{g}r^4h_{\textrm{g}}^{-2}\Omega_\textrm{k}^2,
\end{equation}
with $M_{\rm{p}}$ the mass of the planet, $\Omega_{\rm{k}}$ the Keplerian angular velocity, and $\Sigma_{\rm{g}}$ the gas surface density.

The calculation of the total torque that the disk exerted onto the planets $\Gamma_{\rm{total}}$ is computed according to the planetary mass regime
For planets with masses above 0.01 M$_\oplus$ and below 66 M$_\oplus$ corresponding to a planet–star mass ratio of $q=0.002$ for a 0.1 M$_\odot$ and with $K=\left(\frac{M_{\rm{p}}}{M_\star}\right)^{2} h_{\rm{g}}^{-5} \alpha_{\rm{t}}^{-1} < 10^{4}$, we apply the torque prescription of \citet{Kanagawa2018}, which extends the type~I migration regime. In this case, the total torque is
\begin{equation}
    \Gamma_{\rm{total}}=\frac{\Delta_{\rm{L}}\Gamma_{\rm{L}}+\Delta_{\rm{C}}\Gamma_{\rm{C}}exp(-K/K_{\rm{t}})}{1+0.04K}
    \label{eq:K18}
,\end{equation}
with $\Gamma_{\rm{C}}$ and $\Gamma_{\rm{L}}$ the corotation and Lindblad torques, respectively, for a circular and co-planar planetary orbit (as in \citealt{Paardekooper2010,Paardekooer2011}) and $\Delta_{\rm{C}}$ and $\Delta_{\rm{L}}$ the associated reduction factors for an eccentric and inclined orbit (as in \citealt{Ida2020}). The details of the estimation of the Lindblad and corotation torques, together with the corresponding reduction factors, are described in \citet{Sanchez2022}, where they were implemented for the first time. In its original form, the fit from hydrodynamical simulations made by \citet{Kanagawa2018} does not include the reduction factors, as showed in Eq. \eqref{eq:K18}. However, by treating their prescription as an extension of type~I migration and following the assumptions of \citet{Matsumura2021}, we incorporated these factors into the general fitting formula. The parameter $K_\mathrm{t}$ is associated with the gap depth at which the corotation torque becomes negligible. \citet{Kanagawa2018} showed that for $K \gtrsim 20$ the gap opened by the planet modifies the torque, and therefore adopted $K_\mathrm{t} = 20$. For $K \ll K_\mathrm{t}$ and $0.04K \ll 1$, the total torque reduces to the linear expression, $\Gamma_{\rm{total}}=\Delta_{\rm{L}}\Gamma_{\rm{L}}+\Delta_{\rm{C}}\Gamma_{\rm{C}}$. We note that adopting $K_{\rm t} = 20$ corresponds to the transition mass $M_{\rm tran}$ for partial gap–opening planets proposed by \citet{Kanagawa2018}, assuming a gas surface density reduced by a factor of 0.5. This transition mass is
\begin{equation}
    M_{\mathrm{tran}} = 2.7~ \left( \frac{\alpha_{t}}{0.001} \right)^{0.5} \left( \frac{h_{\rm{g}}}{0.05} \right)^{2.5} \left(\frac{M_\star}{0.1 \mathrm{M_\odot}}\right)~~~~ \mathrm{M_\oplus}.
    \label{eq:/mtrans}
\end{equation}

For planets with masses greater than 66 M$_\oplus$ and values of $K>10^{4}$, we adopted a new prescription for the estimation of the total torque estimation given by Sanchez et al (in review), described as
\begin{equation}
\Gamma_{\rm{total}}/\Gamma_0 = 
\begin{cases}
 ~~a, & \text{if } ~~\Sigma_0 > 10^{-5} \\
  ~~ \frac{b}{q},  & \text{if }~~ \Sigma_0 \le 10^{-5}
\end{cases}
\label{eq:S25}
,\end{equation}
with $a=5\times10^{-4}$ and $b=10^{-5}$, and $\Sigma_0$ the local gas-disk density at the location of the planet $r_0$ in units of $M_\star r_0^{-2}$. 
We emphasize that, in this new regime, planets experience outward migration throughout the disk, in contrast to the previous regime, where migration is predominantly inward except in a narrow region near the inner disk edge (see details on type~I migration in \citealt{Sanchez2022}). We also highlight that this is the first time that such a prescription has been incorporated into an N-body model to study the formation and evolution of giant planets.

We note that all the above equations are evaluated at the semi-major axis of the orbit of the planets.  The remaining timescales $t_{\rm{e}}$ and $t_{\rm{i}}$ are calculated as 
\begin{equation}
    t_\textrm{e} = (1 + 0.04K)\frac{t_\textrm{wave}}{0.78}\left[1+\frac{1}{15}(e_\textrm{rat}^{2}+i_\textrm{rat}^{2})^{3/2}\right],
\label{eq:teIDA}
\end{equation}
\begin{equation}
     t_\textrm{i} = (1 + 0.04K)\frac{t_\textrm{wave}}{0.544}\left[1+\frac{1}{21.5}(e_\textrm{rat}^{2}+i_\textrm{rat}^{2})^{3/2}\right],
\label{eq:tiIDA}
\end{equation}
with $e_\textrm{rat}=e/h_\textrm{g}$ and $i_\textrm{rat}=i/h_\textrm{g}$. Due to the lack of consensus on eccentricity and inclination damping in the type II regime, we follow \citet{Matsumura2021} and adopt the damping timescales by scaling those of type I migration from \citet{Ida2020} with a factor $(1 + 0.04K)$. This ensures that the eccentricity and inclination damping timescales remain shorter than the migration timescale, an essential condition for planets to be resonantly trapped near the inner edge of the disk \citep{Ogihara2010}.
    
As an example, Fig.~\ref{fig:torques} shows the normalized total torques experienced by planets of different masses throughout a 1 Myr disk (see details of the disk model in \citealt{Sanchez2024}), at various distances from a 0.1~M$_\odot$ star, assuming coplanar and circular orbits. The figure illustrates type I migration as described by linear theory \citep{Paardekooper2010,Paardekooer2011,Ida2020}, with a transition at the mass $M_\mathrm{tran}$ (Eq. \eqref{eq:/mtrans}) where type I migration can be extended following \citet{Kanagawa2018}. In this regime, the migration slows down due to a reduction in the total normalized torque. In both cases, we adopt the torque prescription given by Eq. \eqref{eq:K18}. For more massive planets satisfying a planet-to-star mass ratio $q > 0.002$ and $K > 10^{4}$, the normalized torque is calculated according to Eq.~\eqref{eq:S25}. The figure highlights the different regimes of inward and outward migration in each case.

\section{Star--planet tidal interactions}

We adopted the equilibrium tidal model proposed by \citet{Hut1981} and reformulated by \citet{Bolmont2011} to describe the star–planet tidal interactions in systems of planets orbiting brown dwarfs and M dwarfs. This model was originally implemented in the N-body code by \citet{Sanchez2020}. The tidal forces acting on each planet account for the tides raised both by the host star on the planet and by the planet on the star. The acceleration induced on the planets by tidal forces includes terms that affect the evolution of the argument of periastron, $\omega$ ($f_{\rm tide-\omega}$), as well as terms that influence the evolution of the semi-major axis, $a$, and eccentricity, $e$ ($f_{\rm tide-ae}$), as follows:

\begin{equation}
  \textbf{f}_{\textrm{tide}-\omega} = -3\frac{\mu}{r^8}\left[k_{2,\star}\left(\frac{M_\mathrm{p}}{M_\star}\right)R_\star^5 + k_{2,\mathrm{p}}\left(\frac{M_\star}{M_\mathrm{p}}\right)R_\mathrm{p}^5\right] \textbf{r},
\end{equation}

\begin{align*}
\textbf{f}_{\textrm{tide-ae}} = & -3\frac{\mu}{r^{10}} \left[
  \frac{M_{\textrm{p}}}{M_\star} k_{2,\star} \Delta \mathrm{t}_\star R_\star^{5}\left(2\textbf{r}(\textbf{r} \cdot \textbf{v}) + r^{2}(\textbf{r} \times \Omega_\star + \textbf{v})\right)\right]
\end{align*}
\begin{equation}
-3\frac{\mu}{r^{10}} \left[\frac{M_\star}{M_{\textrm{p}}}k_{2,\textrm{p}} \Delta \mathrm{t}_{\textrm{p}} R_\textrm{p}^{5}
       \left(2\textbf{r}(\textbf{r}\cdot\textbf{v}) + r^{2}(\textbf{r} \times \Omega_\textrm{p} + \textbf{v})\right)\right],
\end{equation}

\noindent with $\textbf{r}$ and $\textbf{v}$ the position and velocity vector of each planet, respectively, $\mu=G(M_\star + M_\textrm{p})$, $M_\textrm{p}$ and $R_\textrm{p}$ the mass and radius of the planet, $k_{2,\star}$ and $k_{2,\textrm{p}}$ the potential Love numbers of degree 2 of the star and the planets, and 
$\Delta \mathrm{t}_\star$ and $\Delta \mathrm{t}_\mathrm{p}$ the time-lag model constants for the star and each planet, respectively. The factors
$k_{2,\star}\Delta \mathrm{t}_\star$ and $k_{2,\mathrm{p}}\Delta \mathrm{t}_\mathrm{p}$ are related with the dissipation factors by
\begin{equation}
  k_{2,\mathrm{p}} \Delta \mathrm{t} _\mathrm{p} =  \frac{3R_\mathrm{p}^5\sigma_\mathrm{p}}{2G} \\
   k_{2,\star} \Delta \mathrm{t}_{\star} = \frac{3R_{\star}^5\sigma_{\star}}{2G}
,\end{equation}  
with the dissipation factor for each planet
$\sigma_\mathrm{p}$, 
and the dissipation factor of the host star
$\sigma_\star=2.006\times10^{-53}~\mathrm{k^{-1} m^{-2} s^{-1}}$, the same factor for a M dwarf \citep{Hansen2010}.

We include the evolution of the stellar rotational velocity, $\Omega_\star$, following the prescription proposed by \citet{Bolmont2011}, but neglecting the impact on the planets in its evolution, for simplicity.
We also account for the evolution of the stellar radius, $R_\star$, by fitting the models of \citet{Baraffe2015} for a star with a mass of 0.1~M$_\odot$. We assume that each planet’s rotational velocity, $\Omega_{\rm p}$, corresponds to the pseudo-synchronous state defined by \citet[see further details in \citealt{Sanchez2020}]{Hut1981}. \\

 In this work, we compute the planetary radius assuming a spherical body and adopt three different bulk densities depending on the planetary mass. For planets with $M_{\rm{p}}<3$~M$_\oplus$ we assume a density $\rho=5$~g cm$^{-3}$, for those with $3<M_{\rm{p}}/$M$_\oplus<10$, we use $\rho=2.5$~g cm$^{-3}$; and for planets with $M_{\rm{p}}>10$~M$_\oplus$, we adopt $\rho=1$~g cm$^{-3}$. These values are consistent with typical densities of planets in the Solar System. Moreover, we assume the same tidal dissipation factor for planets with $M_{\rm{p}}<10$~M$_\oplus$, as that of the Earth, $\sigma_\mathrm{p}=8.577\times10^{-43}~\mathrm{k^{-1} m^{-2} s^{-1}}$, estimated by \citep{Neron1997}. For planets $M_{\rm{p}}>10$~M$_\oplus$, we adopt the dissipation factor of Jupiter, $\sigma_\mathrm{p}=7.024\times10^{-52}~\mathrm{k^{-1} m^{-2} s^{-1}}$, following \cite{Hansen2010}.

\end{appendix}

\end{document}